\documentclass[a4paper,11pt]{article}

\usepackage{amsmath}

\usepackage[dvips]{graphicx}

\title{\textbf{Black Body Radiation Shift of the $^{133}$Cs
Hyperfine Transition Frequency}}
\author{Salvatore Micalizio, Aldo Godone, Davide Calonico, Filippo Levi, and Luca
Lorini \\ {\small Istituto Elettrotecnico Nazionale Galileo
Ferraris, Strada delle Cacce 91, 10135 Torino, Italy}}
\begin{document}

\maketitle
\section*{Abstract}
We report the theoretical evaluations of the static scalar
polarizability of the $^{133}$Cs ground state and of the black
body radiation shift induced on the transition frequency between
the two hyperfine levels with $m_{F}=0$. This shift is of
fundamental importance in the evaluation of the accuracy of the
primary frequency standards based on atomic fountains and employed
in the realization of the SI  second in the International Atomic
Time (TAI) scale at the level of $1\times10^{-15}$.

Our computed value for the polarizability is
$\alpha_{0}=(6.600\pm0.016)\times10^{-39} \ \mathrm{Cm^{2}/V}$ in
agreement at the level of $1 \times 10^{-3}$ with recent
theoretical and experimental values. As regards the black body
radiation shift we find for the relative hyperfine transition
frequency $\beta=(-1.49\pm 0.07)\times 10^{-14}$ at $T=300$ K in
agreement with frequency measurements reported by our group and by
Bauch and Schr\"{o}der [Phys. Rev. Lett. \textbf{78}, 622,
(1997)]. This value is lower by $2 \times 10^{-15}$ than that
obtained with measurements based on the dc Stark shift and with
the value commonly accepted up to now. \
\\ PACS number(s): 32.60.+i, 32.10.Dk, 06.30.Ft
\newpage
\section*{I. Introduction}
The evaluation of the accuracy of the cesium primary frequency
standards requires to take into account, among others, the shift
induced on the clock transition $|6 ^{2}S_{1/2}, F=4,m_{F}=0
\rangle \rightarrow |6 ^{2}S_{1/2}, F=3,m_{F}=0 \rangle$ by the
black body radiation (BBR). The definition of the second in the
International System of Units (SI) is in fact based on the above
transition observed in absence of any perturbation and then, in
particular, at $T=0$ K \cite{CGPMresolution}, where $T$ is the
temperature of the environment.

Gallagher and Cook \cite{Gallagher} firstly considered the effects
of BBR on the Rydberg states of the atoms while Itano et al.
\cite{Itano} focused their attention onto the atomic frequency
standards. They reported the following expression for the relative
frequency shift of the $^{133}$Cs clock transition
($\nu_{0}=9192631770$ Hz):
\begin{equation}\label{eq:itano}
\frac{\Delta\nu}{\nu_{0}} =\beta \bigg
(\frac{T}{T_{0}}\bigg)^{4}\bigg[1+\epsilon \bigg(\frac{T}{T_{0}}
\bigg )^{2} \bigg]
\end{equation}

with $\beta=-1.69(4)\times 10^{-14}$ and $\epsilon=1.4 \times
10^{-2}$ when $T_{0}=300$ K. This expression has been adopted till
now to correct all the primary frequency standards for the BBR
shift at an accuracy level of $10^{-15}$ or even better
\cite{steve}.

The BBR induces both non resonant ac Stark and Zeeman shifts of
the atomic transitions. For the ground state $^{2}S_{1/2}$ of the
alkali-metal atoms the Zeeman shift is of the order of $10^{-17}$
\cite{Itano} and is not relevant for the accuracy estimation of
the present frequency standards.

The direct experimental method to evaluate the BBR shift is based
on (\ref{eq:itano}): the frequency shift $\Delta \nu$ is measured
versus the black body temperature \textit{T} and the experimental
points are fitted with (\ref{eq:itano}) whose temperature
dependence follows directly from the Planck radiation law. This
method has been used by Bauch and Schroeder \cite{bauch} employing
a thermal cesium beam and in our group using an atomic fountain
\cite{noi}.

In order to increase the experimental resolution, the BBR shift
may be also evaluated indirectly by measuring the dc Stark shift
coefficient $k$ defined as:
\begin{equation}\label{eq:kappa}
\Delta\nu=kE^{2}
\end{equation}
where $\Delta \nu$ is the frequency shift induced by the static
electric field $E$. Measurements of $k$ have been performed in
this way by Haun and Zacharias \cite{hann} and by Mowat
\cite{Mowat}, using a thermal beam and the Ramsey interrogation
technique, and by Simon et al. \cite{simon} using an atomic
fountain. The BBR shift is then obtained from (\ref{eq:kappa})
assuming \textsl{E}:

\begin{equation}\label{eq:campoe}
E^{2}= \langle \mathcal{E}^{2}(t) \rangle = \frac{4 \sigma
T_{0}^{4}}{\varepsilon_{0} c} \bigg(\frac{T}{T_{0}} \bigg)^{4}
\end{equation}
where $\langle \mathcal{E}^{2}(t) \rangle$ is the mean-squared
electric field of the black body radiation, $\sigma=5.670400(40)
\times 10^{-8} \quad \mathrm{W/m^{2}K^{4}}$ is the
Stefan-Boltzmann constant \cite{codata}, $\varepsilon_{0}$ is the
vacuum permittivity and $c$ is the speed of light in vacuum. In
(\ref{eq:campoe}) the first identity is valid at a level of a few
percent and can be corrected as reported in \cite{Itano}, while
the second identity is obtained directly by the Planck radiation
law (see next section).

The measurement of the scalar static polarizability $\alpha_{0}$
is related to the above method through the expression:
\begin{equation}\label{eq:ploarizzability}
W=-\frac{1}{2}\alpha_{0}E^{2}
\end{equation}
where $W$ is the potential energy of a neutral atom in a static
electric field $E$ of moderate strength. Due to the very low
intensity of the black body field, the hyperpolarizability effect
is negligible \cite{palchikov} and will not be considered
throughout the paper.

The polarizability is obtained from the force acting on the atom
when submitted to a gradient of the electric field:
\begin{equation}\label{eq:forza}
\mathbf{F}=-\nabla W = \alpha_{0}E \nabla E
\end{equation}

This force, in turn, can be measured through the deflection of
thermal beams \cite{Hell, chamberlain}, through E-H gradient
balances \cite{salop, molaff} or through the time of flight of
cooled atoms in a fountain apparatus \cite{maddi,amini}. In
principle, the BBR shift  can be also obtained from
(\ref{eq:campoe}) and (\ref{eq:ploarizzability}) as follows:
\begin{equation}\label{eq:shiftalfa}
\Delta \nu=\frac{1}{h}(W_{\beta}-W_{\alpha})=-\frac{1}{2h}
\big[\alpha_{0}(|\beta \rangle)-\alpha_{0}(|\alpha
\rangle)\big]\langle\mathcal{E}^{2}(t)\rangle
\end{equation}
where $|\beta\rangle = |6 ^{2}S_{1/2}, F=4,m_{F}=0\rangle$ and
$|\alpha\rangle = |6 ^{2}S_{1/2}, F=3,m_{F}=0\rangle$ in the Cs
case; $h$ is the Planck constant.

The indirect experimental technique allows a higher resolution,
being possible to submit the atoms to dc fields of several MV/m,
if compared to the black body field
$\sqrt{\langle\mathcal{E}^{2}(t)\rangle}=832.2$ V/m at \textit{T}
= 300 K used in the direct method. Nevertheless, it is based on
the conceptual assumption $E^{2}=\langle\mathcal{E}^{2}(t)\rangle$
and may not be free from spurious effects as we shall discuss
further on.

>From the theoretical point of view, the evaluation of the BBR
shift requires the computation of the scalar polarizabilities of
the two states defining the clock transition or, in other words,
of the non resonant ac Stark shift. This is typically performed
through perturbation techniques \cite{Landau, Townes}. More
precisely, according to (\ref{eq:shiftalfa}), we have to calcolate
the differential polarizability between the two ground state
hyperfine levels. As pointed out by Feichtner et al.
\cite{Feichtner}, this requires to consider in the theoretical
analysis a basis of modified eigenfunctions which account for the
hyperfine interaction of the ground and $P$ states with other
states of the same quantum numbers $F$ and $m_{F}$. A review of
the theoretical works leading to the evaluation of the ground
state $^{133}$Cs polarizability can be found in \cite{molaff} and
in \cite{amini}.

Due to the basic importance of the BBR shift correction to
evaluate the present primary frequency standard accuracy at the
$10^{-15}$ level and, in the next future at the $10^{-16}$ level,
it is highly desirable to re-examine the relation (\ref{eq:itano})
considering that: \begin{itemize}
\item[(i)]
new and more precise experimental data are now available for the
electric dipole moments and for the transition frequencies of
$^{133}$Cs involved in the theoretical computation of
$\alpha_{0}$;
\item[(ii)]
recent theoretical evaluations of $\alpha_{0}$ \cite{Derevianko}
lead to a significantly lower value than that commonly assumed up
to now \cite{Itano, Mowat};
\item[(iii)]
the two most precise measurements \cite{simon, amini} of the
electrical polarizability do not agree to each other.
\end{itemize}
In this paper we report an ab-initio computation of the BBR shift
for the clock transition which turns out lower by $2 \times
10^{-15}$ at 300 K than the commonly adopted value of
(\ref{eq:itano}). A critical comparison with the theoretical and
experimental values reported in the literature will be also given,
showing the present difficulty to asses the accuracy at the
$10^{-16}$ level for atomic frequency standards operating at T
$\simeq 300$ K.

\section*{II. Basic Theory}
The non-resonant ac Stark shift $\Delta W_{\alpha}^{(k)}$ of the
energy of state $|\alpha\rangle$ induced by the level $|k\rangle$
is given by \cite{Townes}:
\begin{equation}\label{eq:acss}
\Delta W_{\alpha}^{(k)}=\frac{|d_{\alpha k}|^2}{h}
\int_{0}^{+\infty}\frac{\nu_{\alpha}-\nu_{k}}{(\nu_{\alpha}-\nu_{k})^{2}-\nu^{2}}\mathcal{E}_{\nu}^{2}(\nu)d\nu
\end{equation}
where $d_{\alpha k}=e\langle\alpha|\mathbf{e}\cdot \mathbf{r}
|k\rangle$ is the electric dipole moment of the transition
$|k\rangle\rightarrow|\alpha\rangle$, $e$ the electron charge,
$\mathrm{\mathbf{e}}$ the polarization vector of the electric
field, $\mathbf{r}$ the atomic position vector,
$\nu_{\alpha}=W_{\alpha}/h$ and $\nu_{k}=W_{k}/h$ the energy
levels of the states $|\alpha\rangle$ and $|k\rangle$ in frequency
units and $\mathcal{E}_{\nu}^{2}(\nu)$ the mean-squared value of
the electric field at frequency $\nu$. When $|\alpha\rangle$
represents a ground state, we have
$\nu_{\alpha}-\nu_{k}\equiv-\nu_{\alpha k}=-c/\lambda_{\alpha k}$
being $\nu_{\alpha k}$ and $\lambda_{\alpha k}$ the frequency and
the wavelength of the atomic transition $|k\rangle\rightarrow
|\alpha\rangle$. Figure 1 summarizes the above definitions.
\begin{figure}
\begin{center}
\includegraphics[height=200pt]{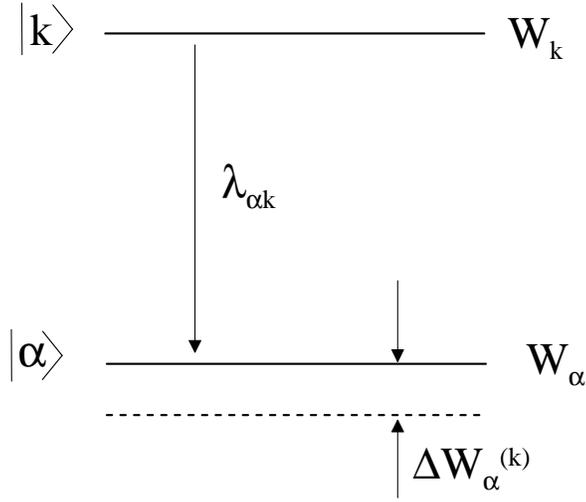}
\caption{Atomic levels considered in (\ref{eq:acss}).}
\label{fig1}
\end{center}
\end{figure}
The electric field per frequency unit $\mathcal{E}_{\nu}(\nu)$ is
related to the spectral energy density $E_{\nu}(\nu, T)$ by the
well known relation:
\begin{equation}\label{eq:campoenergy}
\mathcal{E}_{\nu}^{2}(\nu)d\nu=\frac{1}{\varepsilon_{0}}E_{\nu}(\nu,T)d\nu
\end{equation}
In our case the energy density $E_{\nu}(\nu,T)d\nu$ is given by
the Planck radiation law:
\begin{equation}\label{planck}
E_{\nu}(\nu,T)d\nu=\frac{8 \pi
h\nu^3}{c^{3}}\frac{d\nu}{e^{h\nu/k_{B}T}-1}
\end{equation}
where $k_{B}$ is the Boltzmann constant.

The maximum of $E_{\nu}(\nu,T)$ is reached when (Wien law):
\begin{equation}\label{eq:Wien}
\nu=\nu_{max}=2.821\frac{k_{B}}{h}T
\end{equation}
At \textit{T} = 300 K, relation (\ref{eq:Wien}) gives $\nu_{max}=
17.6$ THz. In the Cs case, the minimum value of $\nu_{\alpha k}$
corresponds to the D$_{1}$ optical transition (335 THz) so that
$\nu_{max} \ll \nu_{\alpha k}$ for all $|k\rangle$ levels coupled
to the state $|\alpha\rangle$ and it is possible to omit the
$\nu^{2}$ term in the denominator of (\ref{eq:acss}) avoiding the
singularity of the integrand at $\nu=\nu_{\alpha k}$. We shall
consider further on the effect of this approximation. The basic
relation (\ref{eq:acss}) can then be written as:
\begin{equation}\label{eq:acss2}
\Delta W_{\alpha}^{(k)}=-\frac{1}{h}\frac{|d_{\alpha
k}|^2}{\nu_{\alpha k}} \langle\mathcal{E}^{2}(t)\rangle
\end{equation}
where the mean-squared electric field
$\langle\mathcal{E}^{2}(t)\rangle$ is given by
\begin{equation}\label{eq:campoenergy2}
\langle\mathcal{E}^{2}(t)\rangle \equiv
\int_{0}^{+\infty}\mathcal{E}_{\nu}^{2}(\nu)d\nu=\frac{4 \sigma
T^4}{\varepsilon_{0}c}
\end{equation}
The relation (\ref{eq:campoenergy2}), obtained through
(\ref{eq:campoenergy}) and (\ref{planck}), is also known as the
Stefan-Boltzmann law; at \textit{T}=300 K it gives
$\langle\mathcal{E}^{2}(t)\rangle=(832.2)^{2}$ (V/m)$^{2}$, as
already reported in the previous section.

Summing over all the excited states $|k\rangle$ coupled to the
level $|\alpha\rangle$, we obtain the total Stark shift for the
state $|\alpha\rangle$:
\begin{equation}\label{eq:acss3}
\Delta W_{\alpha}=-\frac{1}{h}\sum_{k}\frac{|d_{\alpha
k}|^2}{\nu_{\alpha k}} \langle\mathcal{E}^{2}(t)\rangle
\end{equation}
Moreover, taking (\ref{eq:campoe}) and (\ref{eq:ploarizzability})
into account, it is possible to write (\ref{eq:acss3}) also in the
following form:
\begin{equation}\label{eq:acss4}
\Delta W_{\alpha}=-\frac{1}{2}\alpha_{0}
\langle\mathcal{E}^{2}(t)\rangle
\end{equation}
where
\begin{equation}\label{eq:alfa}
\alpha_{0}=\frac{2}{h}\sum_{k}\frac{|d_{\alpha k}|^2}{\nu_{\alpha
k }}
\end{equation}
is the scalar polarizability of the state $|\alpha\rangle$ which
may be measured experimentally.

More precisely, the expression (\ref{eq:alfa}) takes into account
the contribution to the polarizability from the valence excited
states. In principle, the total polarizability includes also the
contribution from the core excited states \cite{dereviankoPRL};
however, this is not important for the clock transition shift
since it is common to both the ground state sublevels.

The BBR shift $\Delta\nu$ of the clock transition
$|\beta\rangle\rightarrow|\alpha\rangle$ we are looking for is:
\begin{equation}\label{deltanu}
\Delta \nu=\frac{1}{h}(\Delta W_{\beta}-\Delta W_{\alpha})
\end{equation}
Introducing relations (\ref{eq:campoenergy2}) and (\ref{eq:acss3})
in (\ref{deltanu}) we finally obtain:
\begin{eqnarray}\label{eq:deltanu2}
\Delta\nu & = &-\frac{4\sigma
T_{0}^4}{h^{2}\varepsilon_{0}c}\bigg(\frac{T}{T_{0}}\bigg)^4
\bigg\{\sum_{k}\frac{|d_{\beta k }|^{2}}{\nu_{\beta
k}}-\sum_{k}\frac{|d_{\alpha k }|^{2}}{\nu_{\alpha k}}\bigg\}
\nonumber\\&=& -\frac{2\sigma
T_{0}^4}{h\varepsilon_{0}c}\bigg(\frac{T}{T_{0}}\bigg)^4
\big\{\alpha_{0}(|\beta\rangle)-\alpha_{0}(|\alpha\rangle) \big\}
\end{eqnarray}
We conclude this theoretical review observing that the dc Stark
coefficient $k$ defined in (\ref{eq:kappa}) can be expressed
through the polarizabilities of the ground state levels (see also
(\ref{eq:shiftalfa})) as:
\begin{equation}\label{eq:kappa2}
k=-\frac{1}{2h}
\big\{\alpha_{0}(|\beta\rangle)-\alpha_{0}(|\alpha\rangle)\big\}=-\frac{8}{7}\frac{\alpha_{10}}{h}
\end{equation}
where $\alpha_{10}$ is the scalar differential polarizability
introduced by Sandars \cite{Sandars}. In the experimental
measurement of $k$ performed with a dc electric field a tensorial
component is also present, which has to be subtracted in order to
take the isotropy of the black-body radiation into account
\cite{Itano,simon}.

\section*{III. Numerical Evaluation}
To evaluate numerically the BBR shift given by (\ref{eq:deltanu2})
we need the matrix elements of the operator $\mathbf{e} \cdot
\mathbf{r}$; writing the atomic position vector $\mathbf{r}$ as an
irreducible rank 1 tensor operator we have:
\begin{equation}\label{eq:prodottoscalare}
\mathbf{e}\cdot\mathbf{r} =
r_{0}\cos\theta-\frac{1}{\sqrt{2}}r_{1}\sin\theta e^{-i
\varphi}+\frac{1}{\sqrt{2}}r_{-1}\sin\theta e^{i \varphi}
\end{equation}
where ${\mathbf{r}}$ is expressed in terms of its spherical
components $r_{0}$, $r_{-1}$ and $r_{1}$; $\theta$ (co-latitude)
and $\varphi$ (azimuth) are the polar angles defining the electric
field direction with respect to the quantization axis ($z$ axis).
The well-known Wigner-Eckart \cite{Edmonds} theorem allows us to
simplify the generic matrix element:
\begin{equation}\label{eq:wignereckart}
\langle\ 6S_{1/2}; F, m_{F}| e r_{q} | n'P_{J'}; F', m'_{F}
\rangle\ = Q(F, m_{F} ; F', m'_{F}; J, J'; q)  \langle\ 6S_{1/2}
|| e \mathbf{r} || n'P_{J'} \rangle\
\end{equation}
where the double bars indicate the dipole reduced matrix element
and \textit{q} is the index labelling the component of
$\mathbf{r}$. In (\ref{eq:wignereckart}) we defined:
\begin{equation}\label{eq:q}
\begin{split}
Q(F, m_{F}; F', m_{F'}; J, J'; q) \equiv
(-1)^{F-m_{F}+F'+J+I+1}\sqrt{(2F+1)(2F'+1)} \\ \times \left(
\begin{array}{ccc}
  F & 1 & F' \\
 -m_{F} & q & m'_{F}
\end{array} \right) \left\{
\begin{array}{ccc}
  F & 1 & F' \\
  J' & I & J
\end{array} \right\}
\end{split}
\end{equation}
where the coefficients in round and in curly brackets are the
3-$j$ and 6-$j$ symbols respectively. The primed quantum numbers
refer to the excited states $|n'P_{J'}\rangle$ coupled to the
ground state by allowed electric dipole transitions.

The polarizability of each of the two ground state levels can then
be written as:
\begin{equation}\label{eq:integralone}
\begin{split}
\alpha_{0}(|F, m_{F}\rangle) = & \frac{2}{4\pi h}\int_{4\pi}
d\Omega \sum_{\substack{n'\geq 6 \\J', F', m'_{F}}}
\frac{|\langle\ 6S_{1/2}||er||n' P_{J'} \rangle
|^{2}}{\nu(6S_{1/2} \rightarrow n'P_{J'}; F')} \cdot \bigg|Q(F,
m_{F}; F',m'_{F}; J, J';q=0) \cos\theta \\&-\frac{1}{\sqrt{2}}Q(F,
m_{F}; F',m'_{F}; J, J';q=1)\sin\theta e^{-i\varphi}\\&+
\frac{1}{\sqrt{2}}Q(F, m_{F}; F',m'_{F}; J, J';q=-1)\sin\theta
e^{i\varphi}\bigg|^2
\end{split}
\end{equation}
In (\ref{eq:integralone}) we perform an integration over the solid
angle $\Omega$ to take the isotropic nature of the BBR into
account, consequently the integral is normalized to $4 \pi$;
$\nu(6S_{1/2} \rightarrow n'P_{J'}; F')$ is the frequency of the
transition indicated in parentheses. According to selection rules,
for fixed values of the quantum numbers only one of the three $Q$
terms is not zero.

Actually, as Feichtner et al. \cite{Feichtner} pointed out, the
$S_{1/2}$ state and the $P$ states wave functions are perturbed by
the effect of their hyperfine interaction with other states with
the same $F$ and $m_{F}$ but with different principal quantum
number $n$. Therefore, the dipole matrix elements should not be
calculated between the unperturbed wave functions but between the
perturbed ones defined as:
\begin{equation}\label{eq:statiperturbati}
\begin{split}
\langle 6 \tilde{S}_{1/2}; F| \equiv \langle 6S_{1/2};
F|+\sum_{n=7}^{\infty} a_{n\frac{1}{2}F}\langle n S_{1/2}; F| \\
\langle 6 \tilde{P}_{J'}; F'| \equiv \langle 6P_{J'};
F'|+\sum_{n=7}^{\infty} b_{n J' F'} \langle nP_{J'}; F'| \\
\langle 7 \tilde{P}_{J'}; F'| \equiv \langle 7P_{J'};
F'|+\sum_{\substack{n=6 \\ n \neq 7}}^{\infty} c_{n J' F'} \langle
nP_{J'}; F'|
\end{split}
\end{equation}
The coefficients $a_{n \frac{1}{2}F}$, $b_{nJF}$ and $c_{nJF}$,
given in \cite{Feichtner}, are here reported for convenience.
\begin{equation}\label{eq:abc}
\begin{split}
&a_{nJF}=\bigg(\frac{\psi_{nS}(0)}{\psi_{6S}(0)}\bigg)
\bigg[\frac{33/2-F(F+1)}{8}\bigg] \bigg[\frac{\nu_{0}}
{\nu(nS_{1/2}\rightarrow 6S_{1/2})}\bigg]
\\
&b_{nJF}=\bigg(\frac{\psi_{nS}(0)}{\psi_{6S}(0)}\bigg)\bigg[\frac{63/4+J(J+1)
-
F(F+1)}{32J(J+1)}\bigg]\bigg[\frac{\nu_{0}}{\nu(nP_{J}\rightarrow
6P_{J})}\bigg]\\
\\
&c_{nJF}=\frac{\psi_{7S}(0)\psi_{nS}(0)}{\big(\psi_{6S}(0)\big)^{2}}\bigg[\frac{63/4+J(J+1)
-F(F+1)}{32J(J+1)}\bigg]\bigg[\frac{\nu_{0}}{\nu(nP_{J}\rightarrow
7P_{J})}\bigg]
\end{split}
\end{equation}
and their numerical values are reported in Tables 1, 2 and 3.

These coefficients are very small (of the order of $10^{-6}$) and
their contribution to the absolute value of the scalar
polarizability is completely negligible. However, they must be
taken into account in order to obtain the correct value of the
differential polarizability we are interested in. \
\begin{table}\label{taba}
\caption{Coefficients $a_{nJF}$ (in units of $10^{-6}$) for
perturbed cesium $6S$ wave function.}
\begin{center}
\begin{tabular}{cccc}
\hline \hline
$n$  & \quad$J\backslash F$  &4 &3\\
\hline
7 &1/2  &-3.37 &+4.33\\
8  &1/2 &-1.63 &+2.08\\
9  &1/2  &-1.04  &+1.34\\ \hline \hline
\end{tabular}
\end{center}
\end{table}

\begin{table}[!]\label{tabb}
\caption{Coefficients $b_{nJF}$ (in units of $10^{-6}$) for
perturbed cesium $6P$ wave function.}
\begin{center}
\begin{tabular}{cccccc}
\hline \hline
$n$  & \quad$J\backslash F$ &5 &4 &3 &2\\
\hline
7 &1/2  & $\backslash$& -1.966 &+2.529 & $\backslash$\\
8  &1/2 &$\backslash$&-0.904 &+1.16 &$\backslash$\\
9  &1/2 &$\backslash$ &-0.567  &+0.729 &$\backslash$\\
7  &3/2 &-1.223 &-0.0583  &+0.874 &+1.573\\
8  &3/2 &-0.560 &-0.026  &+0.400 &+0.721 \\
9  &3/2 &-0.351 &-0.017  &+0.250 &+0.452 \\\hline \hline
\end{tabular}
\end{center}
\end{table}

\begin{table}[!]\label{tabc}
\caption{Coefficients $c_{nJF}$ (in units of $10^{-6}$) for
perturbed cesium $7P$ wave function}
\begin{center}
\begin{tabular}{cccccc}
\hline \hline
$n$  & \quad$J\backslash F$ &5 &4 &3 &2\\
\hline
6 &1/2  & $\backslash$& +1.966 &-2.529 & $\backslash$\\
8  &1/2 &$\backslash$&-1.539 &+1.980 &$\backslash$\\
9  &1/2 &$\backslash$ &-0.741  &+0.953 &$\backslash$\\
6  &3/2 &+1.223 &+0.0583  &-0.874 &-1.573\\
8  &3/2 &-0.957 &-0.0455  &+0.684 &+1.230 \\
9  &3/2 &-0.455 &-0.0217  &+0.325 &+0.585 \\\hline \hline
\end{tabular}
\end{center}
\end{table}

The ground state wave function is more sensitive to the hyperfine
interaction so in this calculation the corresponding sum in
(\ref{eq:statiperturbati}) has been extended from $n=7$ to $n=9$;
the coefficients $b_{nJF}$ and $c_{nJF}$ essentially involve
higher order corrections and  we considered only the first term in
the respective perturbative sums.

As an example, we report in the following equation the expression
of the polarizability of the level $|6S_{1/2}; F=4,
m_{F}=0\rangle$ due \textit{only} to the interaction with the
excited level $|6 P_{1/2}\rangle$:
\begin{equation}\label{eq:integralone2}
\begin{split}
&\alpha_{0}(|F=4, m_{F}=0\rangle) =  \frac{2}{4\pi h}\int_{4\pi}
d\Omega \sum_{ F', m'_{F}}\Bigg[\bigg|Q(4, 0; F',m'_{F}; 1/2,
1/2;q=0) \cos\theta \\&-\frac{1}{\sqrt{2}}Q(4, 0; F',m'_{F}; 1/2,
1/2;q=1)\sin\theta e^{-i\varphi}+ \frac{1}{\sqrt{2}}Q(4, 0;
F',m'_{F}; 1/2, 1/2;q=-1)\sin\theta e^{i\varphi}\bigg|^2 \\
&\times \frac{1}{\nu(6S_{1/2} \rightarrow 6P_{1/2}; F')}
\bigg|\langle\ 6S_{1/2}||er||6 P_{1/2} \rangle\ +
\sum_{n=7}^{9}a_{n\frac{1}{2}4}\langle\ nS_{1/2}||er||6 P_{1/2}
\rangle\ +\\& b_{7\frac{1}{2}F'} \langle\ 6S_{1/2} || er ||
7P_{1/2} \rangle\   + \sum_{n = 7}^{9}a_{n \frac{1}{2} 4} b_{7
\frac{1}{2} F'} \langle\ nS_{1/2}||er||7 P_{1/2} \rangle\
\bigg|^{2} \Bigg]
\end{split}
\end{equation}
For both the ground-state hyperfine levels, we calculated similar
contributions due to the excited states $|6 P_{3/2}\rangle$, $|7
P_{1/2}\rangle$, $|7 P_{3/2}\rangle$, $|8 P_{1/2}\rangle$ and $|8
P_{3/2}\rangle$. In the case of $|6 P_{1/2}\rangle$ and $|6
P_{3/2}\rangle$ (D$_{1}$ and D$_{2}$ line, respectively), we also
included  their hyperfine structure \cite{udem}.

To evaluate numerically the absolute and the differential
polarizability we need to know the reduced dipole matrix elements
and the frequencies of the transitions involved in this
calculation. Several experimental \cite{amini,rafac1,rafac2,young}
and theoretical \cite{amiot,safronova,dzuba,blundell} works have
been devoted to the determination of the dipole matrix elements.
In particular, the measurement of the excited states
$|6P_{1/2}\rangle$ and $|6P_{3/2}\rangle$ lifetimes provides a
direct method to know the reduced matrix elements through the
well-known expression of the spontaneous emission:
\begin{equation}\label{eq:emspont}
\frac{1}{\tau_{J'}}=\frac{\omega_{\alpha k}^{3}}{3
\pi\varepsilon_{0}\hbar c^{3}}\frac{1}{2J'+1}|\langle 6
^{2}S_{1/2}\|er \|6 ^{2}P_{J'} \rangle|^{2}
\end{equation}
being $\tau_{J'}$ the lifetime of the transition
$|k\rangle\rightarrow |\alpha\rangle$ . For D$_{1}$ and D$_{2}$
lines, performing an average of the lifetimes reported in Table 2
of \cite{amini} we obtain: $\tau_{1/2}=(34.86 \pm 0.05)$ ns and
$\tau_{3/2}=(30.44\pm 0.04)$ ns. However, due to the dispersion of
these values we considered an error bar that is twice the
calculated one. For the reduced dipole moments
$\langle6S_{1/2}\|er\|nP_{J}\rangle$ ($n$=7,8),
$\langle7S_{1/2}\|er\|7P_{J}\rangle$ and
$\langle6P_{J}\|er\|7S_{1/2}\rangle$ we perform an average over
the theoretical and experimental values reported in Table 6 of
\cite{safronova}. For the matrix elements
$\langle7S_{1/2}\|er\|8P_{J}\rangle$ we use the values of
\cite{blundell} to whom we attribute an uncertainty of the 1\%.
The signs of the matrix elements are chosen according to the
Feichtner et al. paper. All the values of the dipole matrix
elements used in this numerical evaluation are summarized in Table
4.
\begin{table}\label{momenti}
\caption{Reduced dipole matrix elements of the atomic transitions
used in this calculation.}
\begin{center}
\begin{tabular}{cr}
\hline \hline
Dipole Matrix Element  &Value in $10^{-29} \mathrm{C}\cdot \mathrm{m}$\\
\hline
$\langle6S_{1/2}\|er\|6P_{1/2}\rangle$ &  -3.8174(56)$^{\mathbf{a}}$\\
$\langle6S_{1/2}\|er\|6P_{3/2}\rangle$ & -5.3729(70)$^{\mathbf{a}}$ \\
$\langle6S_{1/2}\|er\|7P_{1/2}\rangle$ &  0.237(4)$^{\mathbf{b}}$\\
$\langle6S_{1/2}\|er\|7P_{3/2}\rangle$ &  0.491(3)$^{\mathbf{b}}$\\
$\langle6P_{1/2}\|er\|7S_{1/2}\rangle$ &  3.59(1)$^{\mathbf{b}}$\\
$\langle6P_{3/2}\|er\|7S_{1/2}\rangle$ & 5.487(8) $^{\mathbf{b}}$\\
$\langle7S_{1/2}\|er\|7P_{1/2}\rangle$ & -8.71(11)$^{\mathbf{b}}$ \\
$\langle7S_{1/2}\|er\|7P_{3/2}\rangle$ & -12.11(1)$^{\mathbf{b}}$\\
$\langle6P_{1/2}\|er\|8S_{1/2}\rangle$ & -0.85(9)$^{\mathbf{d}}$\\
$\langle6P_{3/2}\|er\|8S_{1/2}\rangle$ & -1.23(12)$^{\mathbf{d}}$\\
$\langle6S_{1/2}\|er\|8P_{1/2}\rangle$  & -0.066(2)$^{\mathbf{b}}$ \\
$\langle6S_{1/2}\|er\|8P_{3/2}\rangle$& -0.183(2) $^{\mathbf{b}}$\\
$\langle7S_{1/2}\|er\|8P_{1/2}\rangle$ & 0.776(1)$^{\mathbf{c}}$ \\
$\langle7S_{1/2}\|er\|8P_{3/2}\rangle$ & 1.374(14)$^{\mathbf{c}}$ \\
$\langle7P_{1/2}\|er\|8S_{1/2}\rangle$ & 7.41(70)$^{\mathbf{d}}$ \\
$\langle7P_{3/2}\|er\|8S_{1/2}\rangle$ & 11.5(1)$^{\mathbf{d}}$\\
$\langle6P_{1/2}\|er\|9S_{1/2}\rangle$ & 0.46(5)$^{\mathbf{d}}$\\
$\langle6P_{3/2}\|er\|9S_{1/2}\rangle$ & 0.65(6)$^{\mathbf{d}}$ \\
$\langle7P_{1/2}\|er\|9S_{1/2}\rangle$ & -1.67(17)$^{\mathbf{d}}$\\
$\langle7P_{3/2}\|er\|9S_{1/2}\rangle$ & -2.2(2)$^{\mathbf{d}}$\\
\hline \hline
$^{\mathbf{a}}$ {\small see text $\qquad$}\\
$^{\mathbf{b}}$ {\small from Ref. \cite{safronova}}\\
$^{\mathbf{c}}$ {\small from Ref. \cite{blundell}}\\
$^{\mathbf{d}}$ {\small from Ref. \cite{Feichtner}}
\end{tabular}
\end{center}
\end{table}

In Table 5, we report the values of the atomic transitions
frequencies we use. For the frequencies of D$_{1}$ and D$_{2}$
lines we use the most accurate values presently available
\cite{udem}. For the other transitions we use the values reported
in the Basic Atomic Spectroscopic Database of NIST. These
frequencies are known with an accuracy high enough so that we do
not consider their contribution to the evaluation of the
uncertainty in the final result.

\begin{table}\label{frequenze}
\caption{Frequencies of the atomic transitions used in this
calculation.}
\begin{center}
\begin{tabular}{ll}
\hline \hline
$\qquad\qquad\quad$Transition  & Frequency (THz)\\
\hline
$|6S_{1/2};F=4\rangle \rightarrow |6P_{1/2};F'=3\rangle$ & 335.1113702 \\
$|6S_{1/2};F=4\rangle \rightarrow |6P_{1/2};F'=4\rangle$ & 335.1125378 \\
$|6S_{1/2};F=4\rangle \rightarrow |6P_{3/2};F'=3\rangle$ & 351.7215083 \\
$|6S_{1/2};F=4\rangle \rightarrow |6P_{3/2};F'=4\rangle$ & 351.7217095 \\
$|6S_{1/2};F=4\rangle \rightarrow |6P_{3/2};F'=5\rangle$ & 351.7219605 \\
$|6S_{1/2};F=4\rangle \rightarrow |7P_{1/2}\rangle$      & 652.50476 \\
$|6S_{1/2};F=4\rangle \rightarrow |7P_{3/2}\rangle$      & 657.93238 \\
$|6S_{1/2};F=4\rangle \rightarrow |8P_{1/2}\rangle$      & 770.73660 \\
$|6S_{1/2};F=4\rangle \rightarrow |8P_{3/2}\rangle$      & 773.21409 \\
$|6S_{1/2};F=3\rangle \rightarrow |6P_{1/2};F'=3\rangle$ & 335.1205628 \\
$|6S_{1/2};F=3\rangle \rightarrow |6P_{1/2};F'=4\rangle$ & 335.1217305 \\
$|6S_{1/2};F=3\rangle \rightarrow |6P_{3/2};F'=2\rangle$ & 351.7305497 \\
$|6S_{1/2};F=3\rangle \rightarrow |6P_{3/2};F'=3\rangle$ & 351.7307010 \\
$|6S_{1/2};F=3\rangle \rightarrow |6P_{3/2};F'=4\rangle$ & 351.7309021 \\
$|6S_{1/2};F=3\rangle \rightarrow |7P_{1/2}\rangle$      & 652.51395 \\
$|6S_{1/2};F=3\rangle \rightarrow |7P_{3/2}\rangle$      & 657.94157 \\
$|6S_{1/2};F=3\rangle \rightarrow |8P_{1/2}\rangle$      & 770.74579 \\
$|6S_{1/2};F=3\rangle \rightarrow |8P_{3/2}\rangle$      & 773.22328 \\
\hline \hline
\end{tabular}
\end{center}
\end{table}

As far as the $a_{nJF}$, $b_{nJF}$ and $c_{nJF}$ coefficients are
concerned, we observe that in their definition the knowledge of
the wave function at the origin is required. In turn, the wave
function depends on the potential used to describe the interaction
of the valence electron with the nucleus and with the closed
electron shells \cite{Stone}. In other words, this potential
accounts for the departure from the hydrogen potential. Due to the
difficulty in modelizing this interaction, we made a conservative
estimation, attributing an uncertainty of 10\% to the coefficients
$a_{nJF}$, $b_{nJF}$ and $c_{nJF}$.

Inserting all these quantities and the respective uncertainties in
(\ref{eq:deltanu2}) we obtain:
\begin{equation}\label{eq:risfin}
\frac{\Delta \nu}{\nu_{0}}=(-1.49 \pm 0.07)\times
10^{-14}\bigg(\frac{T}{T_{0}}\bigg)^{4}
\end{equation}
In (\ref{eq:risfin}) the uncertainty is mainly due to the
uncertainty of the coefficients $a_{nJF}$, $b_{nJF}$ and
$c_{nJF}$.

We notice that the result (\ref{eq:risfin}) has been obtained by
means of the integration in (\ref{eq:integralone}) over the solid
angle $\Omega$ which has allowed us to calculate the scalar
polarizability we are interested in; as a check, we obtain the
same value using directly the Sandars formula \cite{Sandars}. If
we estimate the BBR shift by using the polarizability value
computed for a well defined orientation of the electric field with
respect to the quantization axis, the non-zero tensorial part of
the polarizability itself leads to small errors in the coefficient
$\beta$. More precisely, $\beta$ changes of about
$3\times10^{-16}$ in $^{133}$Cs, depending on whether the electric
field is parallel or perpendicular to the \textit{z} axis.

>From (\ref{eq:deltanu2}) and (\ref{eq:kappa2}) we have also for
the Stark shift coefficient:
\begin{equation}\label{eq:kappa3}
k=-(1.97\pm0.09)\times10^{-10} \ \mathrm{Hz/(V/m)}^{2}
\end{equation}
and for the differential polarizability:
\begin{equation}\label{eq:diffpol}
\alpha_{10}=(1.14\pm0.05)\times 10^{-43} \ \mathrm{J/(V/m)}^{2}
\end{equation}
Moreover, our computations give for the scalar polarizability
$\alpha_{0}^{v}$ of the 6$^{2}S_{1/2}$ state of $^{133}$Cs due
only to the valence states:
\begin{equation}\label{eq:alfa0}
\begin{split}
\alpha_{0}^{v} = \quad &(6.341 \pm 0.016)\times 10^{-39} \ \mathrm{Cm^{2}/V} \\
= \quad & (56.99 \pm 0.14)\times 10^{-24} \ \mathrm{cm}^{3} \\
= \quad & (384.9 \pm 0.9) a_{0}^{3}
\end{split}
\end{equation}
In (\ref{eq:alfa0}) the first result is the value expressed in SI
units, the second one in CGS units and the third one in atomic
units, being $a_{0}$ the Bohr radius.

Finally, our evaluation of the corrective term $\epsilon$ is (see
Appendix):
\begin{displaymath}
\epsilon=1.4\times 10^{-2}
\end{displaymath}
in agreement with the value reported in \cite{Itano}.

\section*{IV. Discussion of the results}
We firstly examine the scalar polarizability value reported in
(\ref{eq:alfa0}); even if it does not lead directly to the BBR
shift of the hyperfine transition we are looking for, it
represents anyway an important check of a part of our computations
and provides an independent estimation of this important
parameter.
\begin{table}[!]\label{alfa0}
\caption{Cesium ground state polarizability.}
\begin{center}
\begin{tabular}{cl}
\hline \hline
 $\alpha_{0} (\times10^{-39} \ \mathrm{Cm^{2}/V})$  & $\qquad$Reference\\
\hline
$6.587\pm 0.031$ &  Derevianko et al. \cite{dereviankoPRL} \\
$6.596 \pm 0.013$ &  Derevianko and Porsev \cite{Derevianko} \\
$6.600 \pm 0.016$  & this work \\
$6.611 \pm 0.009$&  Amini et al. \cite{amini} \\
\hline \hline
\end{tabular}
\end{center}
\end{table}
The total scalar polarizability $\alpha_{0}$ must take the core
polarizability $\alpha_{0}^{c}$ into consideration which accounts
for approximately the 4 \% of the total value in the case of
$^{133}$Cs \cite{dereviankoPRL}. Our final result for $\alpha_{0}$
is then:
\begin{equation}\label{eq:alfafinale}
\begin{split}
\alpha_{0}=\alpha_{0}^{v}+\alpha_{0}^{c}=&(6.600\pm 0.016)\times
10^{-39} \ \mathrm{Cm^{2}/V}\\=& (59.32\pm 0.14)\times 10^{-24} \
\mathrm{cm}^{3}
\\=&(400.7 \pm 1.0) a_{0}^{3}
\end{split}
\end{equation}
where $\alpha_{0}^{v}$ is given by (\ref{eq:alfa0}) and
$\alpha_{0}^{c}$ is reported in \cite{safronova}.

In Table 6 we report the data found in literature with an
uncertainty lower than 1 \%.

An agreement at a level of $\pm 1\times 10^{-3}$ can be observed,
well inside the quoted error bars, both between the theoretical
computations and with respect to the experimental value of Amini
and Gould \cite{amini}.

In Figure 2 the result of our computations concerning the BBR
shift parameter $\beta$ is compared with the experimental results
reported  in literature.

Our result agrees with the direct measured values while the
indirect measurements, even if in agreement among them, are higher
by three standard deviations. This discrepancy has to be deeply
examined not only for the correct estimation of the differential
polarizability of the ground state hyperfine transition but also
for its impact into the accuracy evaluation of the primary
frequency standards. First of all, the following points have to be
considered:
\begin{figure}[!]
\begin{center}
\includegraphics[height=230pt]{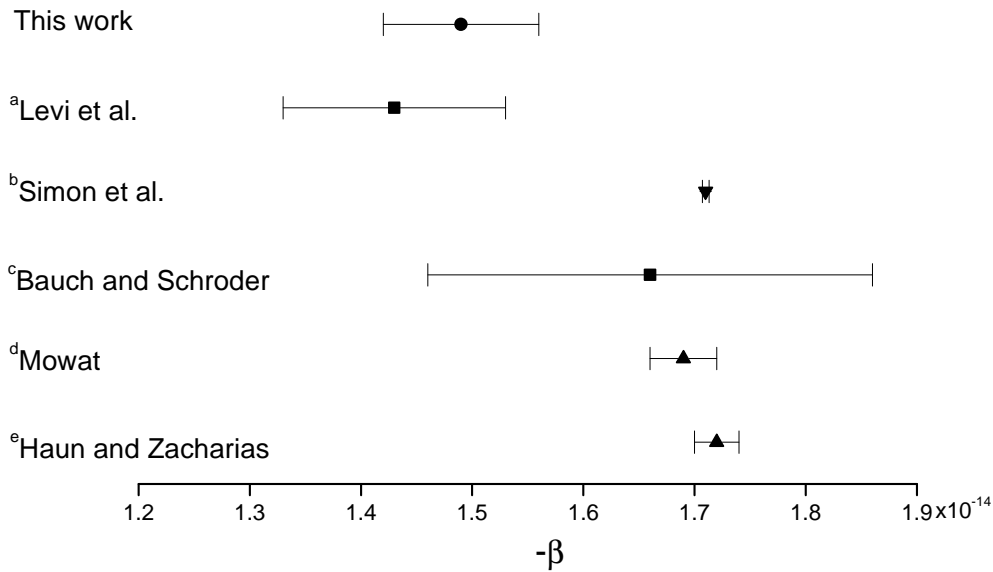}
\caption{BBR shift (parameter $\beta$) of the ground state
$^{133}$Cs hyperfine transition. With the triangle we indicated
the indirect measurements, while with the square the direct ones.
$^{\textrm{a}}$ is Ref. \cite{noi}, $^{\textrm{b}}$ is Ref.
\cite{simon}, $^{\textrm{c}}$ is Ref. \cite{bauch},
$^{\textrm{d}}$ is Ref. \cite{Mowat}, and $^{\textrm{e}}$ is Ref.
\cite{hann}.} \label{fig2}
\end{center}
\end{figure}
\begin{itemize}
\item[(i)]
our theoretical result leads to a polarizability $\alpha_{0}$ of
the $^{2}S_{1/2}$ state in agreement with the most recent and
precise evaluations as shown in Table 6, while the indirect
measurements, considering that the Stark effect is proportional to
the polarizability at the first order, lead to a considerably
higher value for $\alpha_{0}$;
\item[(ii)]
our computations lead to the result of Itano et al. \cite{Itano}
if we use the old values of the transition frequencies
\cite{moore} and of the electric dipole moments \cite{ Feichtner,
Stone}; the latter, in particular, are considerably higher than
the most precise recent values;
\item[(iii)]
our $\beta$ value agrees with the values obtained with the direct
measurement method.
\end{itemize}

Taking the above remarks into account, the following hypothesis
may be considered in order to explain this discrepancy:
\begin{itemize}
\item[(i)]
the assumption of equivalence between ac and dc Stark shifts is
not fully correct, in spite also of the correction term $\epsilon$
introduced by Itano et al. \cite{Itano};
\item[(ii)]
a systematic shift is present in the indirect measurements due to
a physical effect not considered and present in the experiment.
\end{itemize}
Since we have no reason at the moment to believe in the breakdown
of the equivalence between ac and dc Stark shifts, as far as the
second hypothesis is concerned, the following point should be
examined. All the measurements performed with the indirect method
are based on a Ramsey interaction scheme with a Stark field
applied during the free flight of the atoms between the two
interaction regions. The atoms experience then two sudden changes
of the electric field which could perturb the free evolution of
the hyperfine coherence or induce Majorana-type transitions.

The last hypotheses must be considered more a clue than a probe to
explain the discrepancy under question and obviously requires  a
more complete theoretical treatment.

\section*{V. Conclusions}
We have reported in this paper a theoretical evaluation of the
$^{2}S_{1/2}$ ground state polarizability of $^{133}$Cs and the
BBR shift of the ground state hyperfine transition between the
magnetic field independent sublevels. The former is in really good
agreement with recent theoretical and experimental values; the
latter agrees with the direct measurements reported in the
literature and is lower by three standard deviations than the
values obtained from dc Stark measurements. The discrepancy of
$2\times10^{-15}$ at \textit{T}=300 K between the more recent
estimations and the value accepted up to now impacts the accuracy
evaluation of the primary frequency standards at the level of two
or three times their believed standard uncertainties. Moreover,
serious problems may be expected from the BBR shift to asses the
accuracy at the level of $1\times 10^{-16}$ for the future
microwave frequency standards operating at $T\sim 300$ K.

\section*{Appendix}
To avoid the singularity at $\nu=\nu_{\alpha k}$ in section II we
have omitted the $\nu^{2}$ term in the denominator of
(\ref{eq:acss}). This simplification, usually adopted in
literature, is justified by the fact that near room temperature
the energy content of the BBR is peaked at frequencies much lower
than the lowest transition frequency of the atom. An estimation of
the error due to this approximation is provided in \cite{Itano}
with the introduction of the corrective term $\epsilon$ in
(\ref{eq:itano}). Here we re-evaluate this term following also the
method described in \cite{Farley}.

Relations (\ref{eq:acss}), (\ref{eq:campoenergy}) and
(\ref{planck}) give:
\begin{equation}\label{eq:integral}
\Delta W_{\alpha}^{(k)}=-\frac{8 \pi |d_{\alpha k}|^{2}\nu_{\alpha
k}}{\varepsilon_{0}
c^{3}}\bigg(\frac{k_{B}T}{h}\bigg)^{2}\frac{1}{\gamma_{\alpha
k}^{2}}\int_{0}^{+\infty}\frac{1}{1-\frac{x^{2}}{\gamma_{\alpha k
}^{2}}}\cdot\frac{x^{3} dx}{e^{x}-1}
\end{equation}
where $\gamma_{\alpha k}=\frac{h\nu_{\alpha k}}{k_{B}T}\gg 1$ and
$x=\frac{h\nu}{k_{B}T}$. Expanding in (\ref{eq:integral}) the
factor:
\begin{equation}\label{eq:expand}
\frac{1}{1-\frac{x^{2}}{\gamma_{\alpha k}^{2}}} \simeq
1+\frac{x^{2}}{\gamma_{\alpha k}^{2}}+\frac{x^{4}}{\gamma_{\alpha
k}^{4}}+...
\end{equation}
and integrating we obtain:
\begin{equation}\label{eq:bernoulli}
\Delta W_{\alpha}^{(k)}=-\frac{8 \pi |d_{\alpha k}|^{2}\nu_{\alpha
k}}{\varepsilon_{0}
c^{3}}\bigg(\frac{k_{B}T}{h}\bigg)^{2}\sum_{n=1}^{\infty}\bigg(\frac{k_{B}T}{h
\nu_{\alpha k }}\bigg)^{2n}\frac{(2\pi)^{2n+2}|B_{2n+2}|}{4(n+1)}
\end{equation}
being $B_{n}$ the Bernoulli numbers. If we consider  the main
($n=1$) and the first order ($n=2$) terms of (\ref{eq:bernoulli})
and sum over all the valence states, we obtain:
\begin{equation}\label{eq:a3}
\Delta W_{\alpha}=-\frac{4\sigma
T_{0}^{4}}{h\varepsilon_{0}c}\bigg(\frac{T}{T_{0}}\bigg)^{4}\sum_{k}\frac{|d_{\alpha
k}|^{2}}{\nu_{\alpha k}} \bigg\{1+\frac{40
\pi^{2}}{21}\bigg(\frac{k T_{0}}{h \nu_{\alpha
k}}\bigg)^{2}\bigg(\frac{T}{T_{0}}\bigg)^{2}\bigg\}
\end{equation}

We substitute now (\ref{eq:a3}) in (\ref{deltanu}) and after some
algebraic arrangements we obtain:
\begin{equation}\label{deltanu3}
\Delta\nu=-\frac{4\sigma
T_{0}^{4}}{h^{2}\varepsilon_{0}c}\bigg(\frac{T}{T_{0}}\bigg)^{4}\bigg\{\sum_{k}\frac{|d_{\beta
k }|^{2}}{\nu_{\beta k}}-\sum_{k}\frac{|d_{\alpha k
}|^{2}}{\nu_{\alpha
k}}\bigg\}\bigg[1+\epsilon\bigg(\frac{T}{T_{0}}\bigg)^{2}\bigg]
\end{equation}

where
\begin{equation}\label{eq:epsilon}
\epsilon=\frac{40
\pi^{2}}{21}\bigg(\frac{k_{B}T_{0}}{h}\bigg)^{2}\frac{\sum_{k}\frac{|d_{\beta
k }|^{2}}{\nu_{\beta k}^{3}}-\sum_{k}\frac{|d_{\alpha k
}|^{2}}{\nu_{\alpha k}^{3}}}{\sum_{k}\frac{|d_{\beta k
}|^{2}}{\nu_{\beta k}}-\sum_{k}\frac{|d_{\alpha k
}|^{2}}{\nu_{\alpha k}}}
\end{equation}
The main term of (\ref{deltanu3}) coincides with
(\ref{eq:deltanu2}), as expected. As regards the corrective term
$\epsilon$, following the same procedure to evaluate the dipole
matrix elements in the basis of modified eigenfunctions as
reported in section III, we obtain $\epsilon=1.4\times 10^{-2}$,
in agreement with the value reported in \cite{Itano}. This
agreement is not surprising because the recent values of the the
dipole moments affect significantly the main parameter $\beta$,
but much less the corrective term $\epsilon$, as can be observed
in (\ref{eq:epsilon}) where they appear both in the numerator and
in the denominator.

\end{document}